\documentclass[aps, preprint]{revtex4}

\def\be{\begin{equation}}
\def\te{\end{equation}}
\def\ee{\end{equation}}
\def\ba{\begin{eqnarray}}
\def\bea{\begin{eqnarray}}

\def\tea{\end{eqnarray}}
\def\ea{\end{eqnarray}}
\def\eea{\end{eqnarray}}

\def\m{\mu}

\def\D{\Delta}

\newskip\humongous \humongous=0pt plus 1000pt minus 1000pt

\newif\ifdtup

\begin{document}

\title{Gravity and Nonequilibrium Thermodynamics of Classical Matter}
\author{B. L. Hu} \email{blhu@umd.edu. } \affiliation{Maryland Center for Fundamental Physics,
Department of Physics, University of Maryland, College Park, Maryland
20742-4111, USA}
\date{\small October 20, 2010 \\
{\it xx Invited Talk at Mariofest, March 2010, Rosario, Argentina.
Festschrift to appear as an issue of IJMPD}}

\centerline{\it - Dedicated to Professor Mario Castagnino on the
occasion of his 75th birthday.}

\begin{abstract}
Renewed interest in deriving gravity (more precisely, the Einstein equations) from thermodynamics considerations  \cite{Jac95,Jac05} 
is stirred up by a recent proposal that `gravity is an entropic force' \cite{Ver10} (see also \cite{Pad10}). Even though I find the arguments justifying such a claim in this latest proposal rather ad hoc and simplistic compared to the original one I would unreservedly support the call to explore deeper the relation between gravity and thermodynamics, this having the same spirit as my long-held view that {\it general relativity is the hydrodynamic limit \cite{GRhydro,E/QG} of some underlying theories for the microscopic structure of spacetime} -- all these proposals, together with that of \cite{Volovik,Wen}, attest to the emergent nature of gravity \cite{EGrev}.   In this first paper of two we set the modest goal of studying the nonequilibrium thermodynamics of classical matter only, bringing afore some interesting prior results,  without invoking any quantum considerations such as Bekenstein-Hawking entropy, holography or Unruh effect. This is for the sake of understanding the nonequilibrium nature of classical gravity which is at the root of many salient features of black hole physics. One important property of  gravitational systems, from self-gravitating gas to black holes, is their \textit{negative heat capacity}, which is the source of  many out-of-the ordinary dynamical and thermodynamic features such as the non-existence in isolated systems of thermodynamically stable configurations, which actually provides the condition for gravitational stability.  A related property is that, being systems with  \textit{long range interaction}, they are nonextensive and relax extremely slowly towards equilibrium. Here we explore how much of the known features of black hole thermodynamics can be derived from this classical nonequilibrium perspective. A sequel paper will address gravity and nonequilibrium thermodynamics of quantum fields \cite{DICE10}.
\end{abstract}
\maketitle 

{\it Happy Birthday Mario, the philosopher, the king and the philosopher king! -- This was explained in the first few slides of my talk at  Mariofest. In an earlier, different occasion I had compared Mario, in his capacity of an inspiring mentor and a chief architect, in building up an eminent  school of  theoretical physics and astrophysics in Argentina, to my own Ph. D. advisor, the late Professor John Archibald Wheeler, in the U.S.A. The nature of this meeting could perhaps allow me to also relate some of my past experience with Wheeler, and to pay homage to his influence on me in the same capacity as is done here by many young researchers, leaders in their own rights in different fields of physics and astrophysics, with Mario. So please forgive me if you find me delving at times into the past, referring to what I was thinking when I was a graduate student, some 40 years ago,  on certain topics,  some still of current interest. One of these ideas bears on the present theme of gravity in relation to thermodynamics, another on the philosophy I use for understanding it.}

\section{Introduction}

\textit{After Deconstructing Quantum Gravity, - What is Gravity? What is the Quantum?}\\

This is the title of my talk  but this essay addresses only the first question therein, namely, what is gravity, viewed in the light of gravity as an emergent phenomenon, and taking gravity as the thermodynamic limit of some microstructure of spacetime. I endorse this viewpoint as it is on the same footing as my long-held view that general relativity (GR) is the hydrodynamic limit \cite{GRhydro} of some underlying {\it theories for the microscopic structures of spacetime}, the italicized words are what I advocate as the new meaning of quantum gravity (QG). This thermodynamic view of gravity which Jacobson \cite{Jac95,Jac05} started has caught the attention of a broad audience after Verlinde's recent proposal that ``gravity is an entropic force"  \cite{Ver10} (see also \cite{Pad10}). Although I find his reasonings in the derivation of  Newton's second law and the law of universal attraction (drawing on quantum physics in the capacity of Bekenstein entropy \cite{Bek72} and Hawking radiance \cite{Haw74} in black holes, quantum information tied to the holography principle \cite{holography} and the Unruh effect \cite{Unr76} used to relate acceleration to temperature) somewhat overladen and  simplistic  I do support furthering our understanding of the relation between gravity and thermodynamics, examining how Einstein's equations arise in this emergent view  and using the fact that GR is the theory of gravity for our macroscopic world to derive possible thermodynamic principles such as maximum entropy which can be applied to obtain the thermodynamics limit of quantum gravity. In this first paper of two we set the modest goal of studying the nonequilibrium thermodynamics of classical matter only, bringing afore some interesting yet perhaps not so well-known prior results from classical physics to illustrate the nature of gravity, how many salient features of it are common to other interactions and how gravity is different.

To appreciate this new paradigm towards gravity I need to first summarize the key ideas in my view of quantum gravity (or how it is deconstructed) so as to shape the new perspective for understanding classical gravity (this follows  immediately below), and describe what the goal of this paper is, namely, to expound  gravity from a nonequilibrium thermodynamics viewpoint, treating it as a classical system with negative heat capacity (NHC) or as belonging to systems with long range interactions (LRI) and finding out what in gravity is in common with other systems with these properties.  We do this in Sec. 3. In Sec. 4 we will discuss black hole thermodynamics from this  angle, asking if one could derive the Bekenstein entropy formula from a collapsing shell of self gravitating gas, whether the nonextensive character of LRI systems (LRIS) can yield an area- rather than the usual volume- dependence (extensivity) of entropy, and what the failure of the zeroth law of thermodynamics in such systems implies with regard to the  second law of black hole thermodynamics. There is a reason why I want to do it in this methodical way from first principles, namely, to fix the one primary progenitor idea (statistical mechanics of LRI systems) and place the others as secondary or tertiary derived ones. This minimalist attitude I adopt is based on two principles: the Austerity Principle  of Wheeler \cite{Austerity} and my Commonality Principle explained in Sec. 2.   A critique of this recent proposal of Verlinde can be found in the Appendix. In a sequel paper \cite{DICE10} I will discuss gravity with quantum fields from a statistical mechanics perspective and what the gauge/gravity duality implies in this new viewpoint.

\subsection{What is Quantum Gravity?}

Despite the multiplying brands and divergent approaches, from quantum general relativity to loops to string theories and causets \cite{Oriti}, I hope an agreement can be reached soon on a new definition of \textit{quantum gravity (QG) as ``a theory for the microscopic structure of spacetime and matter"} and with it I would propose stop using quantum gravity  in the narrow (and likely misleading) sense of ``quantizing gravity".  The reason is because the only theory of gravity we know and trust is the theory of general relativity (GR) but quantizing GR does not necessarily lead to QG in the sense defined above.  Yet, for 60 yrs this has been the prevailing opinion and practice in the GR community: that a theory for the microscopic structure of spacetime and matter (QG) can be obtained by quantizing general relativity. Most research activities were directed in those decades towards seeking better quantization variables and nicer mathematical formulations. I have had doubts on these programs from the beginning because they all made a \textit{tacit yet unproven assumption}, that quantization of the metric or the connection forms which are macroscopic variables in Einstein's GR theory,  leads to a theory for the microscopic structures of spacetime.  This crucial assumption was never justified, and from what I know in other areas of physics, is generally not true. A different way of thinking with a totally new paradigm is called for. I view general relativity as valid only in the hydrodynamic limit of some (as yet unknown) microscopic theories and as such it is not a fundamental theory, but an emergent one.  The string theory community's view is clear from the beginning: the microscopic theory is string theory. In recent years they announced that gravity is emergent \cite{HorPol,Seiberg}, but how spacetime emerges from string interactions has yet to be shown. Similar claims are made for loop theory \cite{Rovelli,Thiemann}, and therefore, they face the same challenge of showing e.g., the existence of Minkowski spacetime as a stable vacuum at low energies. For the latest progress in causal sets approach see, e.g., \cite{causet}.

Since the main ideas of my view on this subject can be found in earlier papers \cite{GRhydro,E/QG,cosCMP,meso,kinQG,STcond} I will just list some key points and mention their implications as follows.

\paragraph{1. GR as Hydro, metric and connection forms are collective variables}
Direct Implication: {\it Makes no sense to quantize GR.} Doing so will only get the quantized collective degrees of freedom like in phonon physics, not atomic structure.

Note that hydrodynamics here refers to the long wavelength low energy
limit of a micro- theory describing the substructure of spacetime and
matter, not the hydrodynamics of conformal fields in the context of the fluid/gauge duality.

\paragraph{2. Gravity and gauge fields are macro objects, both are emergent from the same micro-theory at its hydrodynamic limit(s).} {\it The existence of a duality or correspondence between them shows that they stand on the same footing.}

Some leading string theorists point at AdS / CFT correspondence as an indication that gravity is emergent. To me this is not hitting the nailhead.   AdS is not emergent from gauge fields nor the converse. {\bf Correspondence is
not emergence.} Gravity and gauge theory could be different manifestations of the same underlying, more fundamental microscopic theory  in different parameter regimes.

\paragraph{3. Macro-structure is largely insensitive to the details of the underlying micro-structures. Many micro-theories can share the same macro-structure.} It is the collective properties of these micro structures which show up at the long wavelength, low energy limit.  In fact one should at the lowest order approximation {\it look first at the commonalities of all competing micro-theories rather than their differences}, namely, their hydrodynamic limits rather than the detailed micro- (string or loop) behavior, in the same sense that magneto-hydrodynamics is based on atoms or ions and magneto-chromo-hydrodynamics based on quarks and gluons share similar features in this limit, namely, hydrodynamics.   The micro-theories could belong to different universality classes similar to the situations in critical phenomena. A classification of these micro-theories in terms of their critical behavior would be very useful in identifying our particular universe. One could then ponder on what other universes could form from different micro-theories under different conditions in their hydrodynamic limits.

\paragraph{4. New tasks of quantum versus emergent gravity defined:} The task of quantum gravity is to induce / discover the micro-structures of spacetime from the known macro-features of our universe while the task of emergent gravity is to explain how these known macro-features of our universe emerge from the assumed micro-theories. I describe this in a recent essay \cite{E/QG}. For the conceptual development of emergent gravity, see \cite{EGrev}.

\section{Commonality Principle}

As remarked in the beginning though sharing the same view that gravity is emergent and GR is the thermodynamic and hydrodynamic limits of some theories for the microscopic structure of spacetime (quantum gravity) I find the arguments backing up the proposal of gravity as entropic force \cite{Ver10} overly simplistic and contrived. As the proposer acknowledged, all the ingredients used are known before, what is different here is just placing a different emphasis on which principles appear to be more fundamental in nature.  Since there is no new results reported compared to the original proposal \cite{Jac05} it boils down to a comparison of the practicing principles and the intellectual attitudes taken in the different approaches. The criterion for deciding what concepts are regarded as primary and fundamental versus what others are secondary and derived is a philosophical issue. Thus foremost we need to recognize differences at the philosophical level -- after all, this is how a viewpoint is formed.

We prefer to evoke the minimal amount of thematic material and work with the lowest common denominator which gravity shares with other physical systems or forces in nature. This philosophy encapsulated in the so-called ``Austerity Principle" or ``Commonality Principle" explained below is in  stark contrast to the ``mix and fix" or ``chop-sui" approach \footnote{`Chop-Sui' is a nondescript Chinese dish invented a century ago in the American Chinatown ghettos
which can neither be found nor even be heard of anywhere in China.
Literally it means ``cut the miscellanies and mix them up".}
to formulating new ideas which seems to be in vogue these days.

`Common' here connotes at least the following three senses: a) that it is not so unusual, particular
or outstanding: It conveys the intent of \textit{seeking a more basic level of understanding}, and in so doing recognize that some new alleged discoveries can in essence be more commonplace than when they are first conjured. b) Common in the sense of common denominator: \textit{seeking a common ground  with other apparently unrelated phenomena}, so as to be able to see their universality.  3)  Common as in commoners: Principles for Commoners. The philosophy behind Wheeler's \textit{``poor man's way" of looking at the essence of matter or getting to the heart of a problem} smacks of this: Find a way that even a poor man can come to appreciate what is valuable in what we want to say.

In applying the Commonality Principle we can usually deconstruct what seems too complex or demystify what appears so intriguing.  Only by performing this screening using these criteria,  by first finding out what characteristic features are in common with other physical phenomena, can we identify what is truly unique and special about a new discovery, in the present case, a new way of understanding gravity in contra- distinction to other forces.
The price one pays for maintaining a self-presumed special stance or unique status is, as we know from the difference between aristocrats and commoners, that despite their imposing appearances or staged elegance they often don't get much done in what really matters, because they are restricted by their own perceived superiority and artificial distinction.

There are many examples we can think of in the history of development of ideas in physics, some more recent examples are: Calabi-Yau space as containing the fermion spectrum in the unification scheme between spacetime and particles. At its inception people thought there was only one such space which matches exactly what we observe in Nature, but of course it was later found that there are millions. Likewise for the quasi-classical domain after decoherence of a quantum system. In the beginning people thought that it is enough to just perform one set of projections in the decoherence functional and then the physical world appears. But later it was found to have infinitely many. String theory is a deep theory but so far rather remote from observation, but the AdS/CFT correspondence \cite{AdSCFT} which was inspired by it took on a thriving life of its own. Yet physics is not quite represented by supersymmetric Yang-Mills theory at the boundary of our universe, so I would take it as  \textit{suggestive} of an interesting and important relation rather than taking all the results of AdS and CFT calculations \textit{literally}.  I believe the gauge/gravity duality \cite{HorPol} which is a representation of this correspondence will produce more useful results because it relates the two well-established and well-understood theories  which form the cornerstones of theoretical physics, in today's familiar low energy rather than in the lofty Planck energy domain.

As a familiar example in gravitation theory, recall Unruh's way of understanding Hawking radiation from
accelerated detectors (1976) \cite{Unr76} or from his fluid model (1981) \cite{Unr81}. The essential physics in
black holes or what special effects may originate from the event horizons can be
understood through more commonplace and reachable physical phenomena. I share the same philosophical perspective.
At the time, the prevailing view of the GR community is that Hawking and Unruh effects are due to the existence of an event horizon which is a global property of spacetime. This is absolutely correct. But if one deviates slightly from that setup and  asks the following question:  \textit{Is there radiation when the detector is accelerating but
not uniformly?}  1) Mathematically, there is no event horizon for such state of motion and the answer should be NO, if we place the presence of an event horizon  as a criterion for the existence of radiation.
However, 2) physically, this is a common practice (of say, passing a slow driver, going from uniform
velocity to uniform acceleration then back to uniform velocity) and a continuous extension of the uniform acceleration case. So one would expect that there is radiation, but not in a thermal form.  The traditional way of using global concepts is too rigid for this problem. A broader way of thinking with a new method is better suited.  We did a calculation using stochastic field theory \cite{RHK} for finite-time uniform accelerated detector  interacting with
a quantum field and found that  2) is the correct answer. This method enables us to calculate the spectrum of radiation from any arbitrary trajectory, no temperature concept is needed because the system is under nonequilibrium condition.


\section{Gravity as nonequilibrium thermodynamics of classical matter}

\subsection{An aside: Some old yet still stimulating ideas}

In the year 1969-70 when I began my doctoral work with Wheeler four
ideas crossed my path and stayed on my mind ever since: 1) Sakharov's
one page 1967 paper about metric elasticity \cite{Sak} which later led to
induced gravity and inspired my view that GR is a hydrodynamic
theory. 2) Wheeler's riddle-like ``boundary of a boundary is zero" motto, which is explained in Chapter 15 of MTW \cite{MTW} and years later further developed into his Austerity Principle \cite{Austerity} 3) Lynden-Bell's paper with Wood \cite{LynWoo68} on the negative heat capacity of gravitating systems which I will expound
further with you in this paper and 4) spectral decomposition of the
Laplacian. Wheeler introduced the work of the mathematician M. Berger
to me, wanting me to use it to see if there is some special state in
the mixmaster universe similar to the magic number in the nuclear
collective model of Hill and Wheeler, Bohr and Mottleson.  All four ideas share one common feature on this curious and uninitiated young mind: unconventional and oblique, yet elegant and appealing; deceptively simple in appearance but probably loaded with deeper meanings. Strange things usually stay longer in one's memory.

Here's a little background: For 1) at that time the big thing in GR
was the  singularity theorem of Penrose, Hawking and
Geroch. Every serious student in general relativity needs to be
conversant in this vigorous mathematical enterprise called global
analysis. So where should one place this idea of Sakharov, fallen
from the blue, totally out of line?  For 2) what Wheeler wanted to convey is that almost all important equations in theoretical physics from QED to QCD to perhaps QGD (quantum gravi-dynamics), those which embody the Bianchi identity, seem to convey something rather obvious, if not `vacuous', because that identity basically states that the boundary of a boundary is zero. This identity signifies the conservation of the `moments of rotation' of a geometric object which corresponds to the conservation of energy-momentum tensor of matter by way of the Einstein equations.  Wheeler later developed this idea further and summarized his understanding in the so-called `Austerity Principle' (essentially saying, if I may use a Wheeleresq expression: that `nothing' is `everything')  This conservation law is what inspired me later to finally affirm that general relativity is geometro-hydrodynamics \cite{GRhydro} (there were earlier hints I noted from cosmology as `condensed matter' and semiclassical gravity as mesophysics \cite{meso}), because, just like hydrodynamics is the long wavelength, low energy limit of the underlying microscopic theories of matter, a geometric (manifold) description of spacetime structure with its dynamics described by the theory of general relativity is meaningful only in the hydrodynamic regime of a microscopic theory of spacetime, which is the proper definition of quantum gravity. This is further corroborated by the findings at that time by Hartle, Laflamme and Marolf \cite{HarLafMar} that a quasi-classical domain of the hydrodynamic variables can exist after decoherence projections because there exist conservation laws that these variables obey.

For 3) we learned that heat capacity for ordinary matter is always positive, it is the condition for the
stability of the canonical ensemble. (We have to know this to pass the
gruesome yet empowering General Exam.)  Now this English gentleman
astronomer is telling us despite everything is coupled
through gravity, gravitational force behaves opposite to almost
everything. I wanted to find out more about this idea, it appears to be the overriding
property of all gravitational phenomena, from self-gravitating gas to
black holes. I didn't get to it though, because I need to study
quantum field theory in curved spacetime! So I had to put this quest
on hold, pretty much until now.

4) is engrained in ``Can you hear the shape of a drum?" I  started reading the series of papers
by Balian and Bloch, where they tried to apply this formula to nuclear
properties. What stuck on my mind was that this is one neat way to
connect the big with the small: the geometry and topology of space
from the eigenvalues  of the invariant operators defined on it. The more I
studied this the more I saw its beauty, but got totally frustrated
because this formula applies only to Riemann and not to
pseudo-Riemann spaces, but I need to construct a formula for the latter to be able to
say something about cosmology. If I were a bit less ambitious and
complacent with spacetimes with Euclidean sections I could have gone
down the route of discovering the zeta function regularization method and
identified the `corner' term in that
expansion as the $a_2$ HamideW coefficient which enters in the conformal anomaly!
(I learned later that great advisors always push their
students beyond the limit, not only of their own ability, but also
beyond what is commonly perceived as acceptable at the frontier of research). This last point is
very powerful -- I saw a connection between geometry-topology and
statistical mechanics, because it provides a microscopic description for the global properties of
spacetime, in terms of the level density of an invariant operator. I
will return to this point in my sequel essay \cite{DICE10}.

\subsection{Old physics: Gravitational systems have negative heat capacity}

Hereafter we will focus just on idea 3). To me the best introduction is still Lynden-Bell's \cite{LynBelPhysica}.
The story began with
Antonsov's 1962 `gravothermal catastrophe' \cite{Ant}: When he analyzed the
thermodynamics of a system of $N$ particles of energy $E$ in a spherical
box of radius $R$ he showed that there is no global maximum entropy
state, meaning that the system is unstable thermodynamically. A local (meta-stable)
maximum entropy state exists when $R$ is not too large.
When $R$ is increased the density of the gas at the edge drops compared
to that at the center. When this ratio decreases below 1/709 this
meta-stable state disappears and the system becomes thermodynamically
unstable. Lynden-Bell and Wood \cite{LynWoo68} offered the explanation that this is because gravitating systems
possess negative heat capacity. Relative stability of stars is made
possible by the balance of another force which provides positive heat
capacity such as the main sequence stars fueled by thermonuclear reaction at the
core. But the outer part of the star, like a self-gravitating gas,
still has negative heat capacity.  Energy released from the core is
absorbed by its outer region expanding and cooling. Thus, according to Posch
and Thirring \cite{PosThi05}, it is the thermal instability
of gravitating systems which extinguishes the nuclear fire and keeps
the star stable. Yet, as we all know, there is an end to nuclear
reaction at which point gravity wins and after a supernova explosion
the core turns into a neutron star or black hole, depending on the
initial mass of the system.  For higher mass systems beyond the
Chandrasekhar limit the degenerate Fermi gas equation of state is too
soft to resist gravity and this `gravothermal catastrophe' leads to
black holes.

All this about the thermodynamics of gravitating systems and
gravitational collapse should be familiar to the gravity and
astrophysics community. For a review, see e.g. \cite{Pad}


The thermodynamical properties of gravitating systems are different from  what we
usually encounter for ordinary matter, where the use of canonical (CE)
ensemble (systems in contact with a large heat reservoir) for their
description is justified. For gravitational systems as well as some
other kinds named below, one needs to revert to the use of
microcanonical (MC) ensembles (for isolated systems at a definite energy
$E$.) Thirring in 1970 \cite{Thi70} showed that two systems with
negative heat capacity (NHC) cannot be in thermal equilibrium. Thus it is
impossible to find an equilibrium canonical ensemble for the combined
system. This may look strange because our usual notion tells us that if one
system A is in equilibrium with another B and if a third one C is
found to be in thermal equilibrium with B, then C should also be in
equilibrium with A. This is canonized as the Zeroth Law of
Thermodynamics. This law implies the existence of an intensive
quantity, the temperature. Notice that the validity of the Zeroth law
has an implicit presupposition, that all systems concerned have positive
heat capacity and have properties adapt to a canonical ensemble description.
These more commonly encountered systems where our notions of
thermodynamics are largely derived from are usually large systems
with short range interactions.  This is not the case for gravity.
\textit{Gravitational systems are nonextensive.} As discovered later,
gravity is not alone in this regard: systems with long range
interactions or small clusters -- small compared to the interaction
range -- have similar thermodynamic properties.

Whether the Zeroth Law is obeyed by systems with NHC is an
interesting issue. Even though we know these systems are
thermodynamically unstable when they interact with their surroundings
and anomalous behavior may ensue, it is not obvious that this
violates the Zeroth Law, because while there is coupling, heat
exchange is allowed and one cannot rule out the possibility that the total system may
become canonical. This is what Ramirez-Hernandez et al \cite{RLL0TD}
set forth to prove, using a small system of rotors in two
dimensions treated both in the MC and in the canonical ensembles.
They confirmed what Thirring proved earlier, that indeed the Zeroth
Law conjured for canonical ensembles does not apply to systems with NHC.
When two identical subsystems of NHC with the same intensive parameter (temperature defined in the MC sense)
are thermally coupled they undergo a process in which the total
entropy increases irreversibly. The intensive parameters of the two
subsystems remain equal but that of the combined system is different
from either subsystem. The two subsystems cannot maintain stable thermal equilibrium.

Knowing that black holes are systems with NHC this result of entropy increase
when applied to black holes becomes a statement of the second law of
BH thermodynamics. If we follow Bekenstein in assigning an entropy to
the black hole  as proportional to its area, then when two
black holes are brought in contact with each other the combined system would see an entropy increase, which would signify an increase in the area of the merged black hole. The association of the area of a
black hole with entropy stems from a different physics, perhaps best
understood from the entanglement entropy viewpoint, to be discussed
further in my sequel essay.  But as far as the entropy increase in
the combined system is concerned, it can be understood as a consequence of both
subsystems having negative heat capacity. We will continue to
explore the thermodynamical properties of gravitating systems with and
without an event horizon, to understand better the role played by the
event horizon over and above their being systems of NHC.

\subsection{Features of gravity common to systems with long range
interactions (LRIS)}

Let's go a step further to seek the commonalities of black holes with other systems.
Negative heat capacity is not unique to gravity. For systems with long-range interactions (LRIS), the two-body potential decays at large distances as  $V(r) \sim 1/ r^a$ with $a \le d$, where $d$
is the space dimension. Examples are: gravitational systems,
two-dimensional hydrodynamics, two-dimensional elasticity, charged
and dipolar systems. Their common properties are:

1.  \textit{Nonextensive / Non-additive}: In normal systems if one holds the
intensive variables (temperature, pressure, chemical potential)
fixed, then the extensive variables (energy, entropy) will increase
in proportion to an increase in the size of the system. This is not
true for LIR systems. Their extensive variables are intrinsically
non-additive: sum of the energies or entropies of subsystems is not
the same as the energy or entropy of the whole system.

2. \textit{Inequivalence of Ensembles}, i.e., for the thermodynamics of such
systems, results obtained from the microcanonical and canonical
ensembles are different.   This inequivalence implies
that specific heat can be negative in the microcanonical ensemble,
and temperature jumps can appear at microcanonical first order phase
transitions. For a discussion of microcanonical thermodynamics, see, e.g.,
\cite{Gross_mc}. For different ensemble treatments of black hole
thermodynamics, see the work of Braden, Brown and York
\cite{BBY}.


3. For LRIS the space of accessible macroscopic thermodynamic parameters might be non-convex.
The lack of convexity allows us to easily spot regions
of parameter space where \textit{ergodicity may be broken}.

4. LRIS also display an extremely slow relaxation towards thermodynamic
equilibrium and \textit{convergence towards quasi-stationary states}.
This includes the famous class of glassy systems, which has been
compared to the thermodynamics of black holes \cite{NiePRL98}
(background temperature referred to there may be inappropriate \cite{Siv}).
See \cite{Campa} for a review on the statistical mechanics and out-of equilibrium
dynamics of systems with long-range interactions.



\subsection{New physics: Quantum Processes in Black holes and holography}

Black hole thermodynamics is often viewed as holding the key to understanding the intersection, if not the union, of
gravitation, quantum mechanics and thermodynamics.  Bekenstein's
identification of black hole surface area as entropy suggests a
strong connection between gravitation and thermodynamics, while Hawking's discovery that black holes emit thermal  radiation takes this to a higher level, incorporating quantum field effects in black hole thermodynamics.

Black holes also have negative heat capacity, no different from a
gravitating gas.   In the 80s there were discussions of this aspect in
the works of Page, Penrose, Smolin,  Sorkin  and others. It is well
known that a black hole in an asymptotically flat spacetime is
unstable. To keep it in quasi-equilibrium we need to place a black
hole in a box (of radius smaller than 3M) or in anti- de Sitter space
where the curvature at infinity acts like a confining wall. Hawking
and Page \cite{HawPag} showed that in an AdS space filled with thermal radiation
the system can undergo a phase transition corresponding to the
formation of a black hole. With the use of the AdS/CFT correspondence
\cite{AdSCFT}, Witten \cite{Witten} showed that this transition would
correspond to the deconfinement transition in QCD. These modern
tenets of holography principle \cite{holography} and gauge/gravity
duality \cite{HorPol} have catapulted black holes in AdS space into special
prominence pertaining to both particle physics and gravitation theory. A lucid
description of the thermodynamics of AdS black holes can be found in
Hemming and Thorlacius \cite{HemTho}.

Why do I pull out this 40 year-old topic of negative heat capacity
for gravitating systems and place it next to the new  fanciful
developments of the last 10 years? Well, here is where the Austerity and
Commonality Principles are at work. What I advocated in our thought
process in relation to some recent hype is, figuratively speaking, to dig deeper at the roots, not just picking the fruits. The gravitational features of systems of interest involved in these new
developments after 1970 -- from black holes entropy to Hawking radiation
to AdS/CFT -- are still largely governed by the thermodynamics of
gravitating systems, not much different from what the simple classical
model of $N$ gravitating particles used 40 years ago by Lynden-Bell and Thirring for understanding stellar evolution. 

Let me illustrate this by examining several aspects of black
hole (BH) physics: I'll divide these issues into two groups, the first on thermo-
statics, such as  area law scaling, phase transition and
meta-stable states, the second on thermo-dynamics, such as BH
formation, BH entropy from collapsing shells and BH nonextensivity.

\section{Black Hole thermodynamics from gravity as systems-with-long-range-interaction viewpoint}

\subsection{Are there substantive differences between a black hole and a
self-gravitating gas in their thermodynamics?}

After understanding some salient features of gravitational systems
from their root up, namely, as systems with negative heat capacity, let us now
consider black holes and ask the question: What is the difference
between a black hole and a self-gravitating gas in their
thermodynamical behavior? One clear distinction is that the red-shift near a
black hole event horizon goes to infinity. This gravitational red-shift factor is from the $g_{00}$ component
of the metric but one can obtain it from special relativity and avoid general relativity all together, in the
same spirit as how Taylor and Wheeler treated the black hole in their book \cite{TayWhe}. Thus keep only
classical gravity and special relativity (or the equivalence
principle, as in \cite{OppSpin}), but no GR or quantum physics.

We will start with the same system of self-gravitating gas used by
Lynden-Bell but consider the full range of total mass to allow for
the formation of a black hole. Let us see if we can obtain
Bekenstein's black hole entropy by considering the collapse of a
self-gravitating gas. Can we see the BH entropy expression emerging from
that of the original gas at the formation of a black hole? This problem was treated by Pretorius, Vollick and Israel \cite{PVI98} 
who calculated the entropy of a thin spherical shell that
contracts reversibly from infinity down to its event horizon and
found that, for a broad class of equations of state, the entropy of a
non-extremal shell is one-quarter of its area in the black hole
limit. (A massive BH was considered to avoid back-reaction effects.)
From this the authors gave an operational definition for the entropy of a
black hole as  the equilibrium thermodynamic entropy that would be
stored in the material which gathers to form the black hole, if all
of this material were compressed into a thin layer near its
gravitational radius. In their derivation they allow for the presence of a
quantum field which imparts a Boulware stress-energy in addition to
the classical stress-energy outside the shell.
To maintain reversibility, the shell must be in equilibrium with the
acceleration radiation seen by observers on the shell.  To maintain
thermal equilibrium they need to draw on a source of energy at
infinity to `top up' the Boulware stress-energy to an appropriate
thermal environment.

This is very nice, yet I would have liked to see a derivation of the black hole entropy-area
law without bringing in any quantum field because I believe this law
fundamentally describes the macroscopic statistical mechanics property of a
classical system with LRI rather than a quantum field-theoretic
property which pertains more to the system's microscopic features.
This wish of mine seems to be fulfilled in an evocative work of
Oppenheim \cite{OppArea} 
where he showed that the scaling laws of the thermodynamical
quantities in the system he studied are identical to those of a black
hole,  even though the system does not possess an event horizon. His
system consists of $n$ densely packed shells supporting itself in its
own gravitational field configuration. Generalization to many shells
extending the work of Pretorius et al allows the system to be compressed to a
size close to its own event horizon. For a system without
gravitational interaction the system's entropy is proportional to its
volume, no surprise.  As weak gravitational interaction is
introduced, the system's entropy acquires a correction term,
beginning to deviate from the familiar volume-scaling properties. In the
limit that the system is about to form a black hole, its entropy is
proportional to its area for a large class of equations of state. The
scaling laws of the system's temperature and energy are also
identical to those of a black hole despite the fact that no horizon
is present. The entropy is found to be proportional to the logarithm
of the number of micro-states of this system. The temperature, which
is usually considered to be independent of the size of the system, is
now inversely proportional to the mass of the system. Oppenheim
concludes that many of the peculiar properties of black hole
thermodynamics are the result of gravitational self-interaction
rather than the presence of a horizon.

A related study of interest in purely classical gravity is by
\cite{OliDam} 
on the dynamics of nonspherical gravitational collapse of a
bound source allowing for mass loss through  gravitational wave
emission. The exterior spacetime these authors used is the class of
Robinson-Trautmann (RT) metric, which contains the simplest
axisymmetric known solutions of Einstein's vacuum field equations
representing an isolated gravitational
radiating system. For sufficiently smooth initial data, the RT metric
converges asymptotically to the Schwarzschild black hole. Physically
the bounded source suffers mass loss by gravitational wave emission
until a Schwarzschild black hole is formed. The authors were
interested in the relation between the fraction of mass radiated away
to the final mass of the bounded source in the nonlinear dynamics
described by Einstein's equations. They found that this does not
depend on the particular form of the initial data families, but
solely on the value of the initial mass, and it satisfies the
distribution law for the (Tsallis) nonextensive statistics. This last
point brings home the properties associated with LRI systems. Another interesting aspect this work on
the critical phenomena of gravitational collapse \cite{GravColcridyn}
brings out is the implications for black hole entropy. As we have
seen above the black hole entropy is not an extensive quantity, being
proportional to its area but not its volume. This is attributed to
the long range interaction nature of the gravitational force.  If
we extrapolate the  results of Pretorius et al and Oppenheim to this
situation, and use the area of the parameter space for the initial
data of the bounded source as a measure of the system's  entropy, the
area of the formed black hole turns out to be proportional to the
black hole entropy. What these authors found is that the final area
is always greater than the initial area.
They asserted that it might be useful to understand the initial data
as an out-of-equilibrium thermodynamical state which evolves towards
the equilibrium state identified as the Schwarzschild black hole.
They also conjectured that, once a given initial data is specified,
the total mass extracted should reflect the nonextensive character of
the black hole entropy. This harks back to a striking feature of all
systems with LRI \cite{Campa}, the \textit{convergence towards
quasi-stationary states}.

I might add two bits of related information here: The mass lost in the form of
gravitational waves is measured by the Weyl curvature, the square of
which enters in Penrose's definition of gravitational entropy
\cite{Penrose} which he introduced to discuss the issue of why the universe
started at such a low entropy state. I augmented this classical
picture by  including quantum matter field in the total entropy
budget and posited that particle creation from the dynamics of
spacetimes (anisotropic and inhomogeneous, with high gravitational
entropy) increases the entropy of matter at the expense of
gravitational entropy (smoothing of the universe) \cite{Hu81}. This
portion of the gravitational energy and entropy corresponding to
gravitational waves (Weyl) is used recently by Chirco and Liberati \cite{Liberati} to fill in
some missing content in  Jacobson's nonequilibrium thermodynamics
formulation of Einstein equation \cite{Jac95}, an extension of his earlier
equilibrium considerations \cite{Jac05}.

In conclusion, let me mention some interesting analog model studies
of such systems.   It is difficult to perform gravitational
experiments in an earth-bound lab on account of the gravitational system's having
negative heat capacity. (One cannot appeal to electric analogs
because for an overall charge-neutral system Lebowitz and Lieb
\cite{LebLie} showed that for Coulomb systems the microcanonical and
canonical ensembles are equivalent.) As we have seen before, a system
with negative specific heat is thermodynamically unstable, namely, if
it is placed in thermal contact with a second system, even a slight
random fluctuation in energy will lower its temperature and increase
the energy transfer further. Nevertheless, one can do modeling and
simulation with clever setups.  Posch and W. Thirring \cite{PosThi06}
demonstrated these microcanonical features with a simple mechanical
model of interacting classical gas particles in a specially confined
domain subject to gravitation. The endgame scenario is that most of
the gas particles are cooled and collect in the lowest part of the
container, where the energy is carried away by a few remaining
particles.

Many thermodynamic properties of gravity are due to gravitational
interaction being long-ranged. Such properties are also
found in small systems. Oppenheim \cite{OppSpin}  
presented a lattice model with long-range spin-spin coupling and
showed that the system's temperature and entropy have many properties
which are found in black holes.  Analog models which can be tested
experimentally or simulated theoretically are useful not only to
pinpoint the physical origin of the salient features in the
thermodynamics of gravitational systems, as is the emphasis of this
essay,  but once such features are identified they can be used to
carry out simulations which are otherwise difficult to perform for
gravitational systems.

\section{Discussions}

Two most familiar models of emergent theories for an average physicist like me are probably hydrodynamics and thermodynamics which describe the robust or stable macroscopic manifestations of a  microscopic theory (or many).  Hydrodynamics is robust because of the existence of conservation laws for the collective or hydrodynamic variables. Thermodynamics describes stationary configurations under equilibrium conditions determined by the maximal entropy principle. In my search for understanding the emergent behavior traversing the familiar macroscopic world into the unknown microscopic structures of spacetime (quantum gravity) I have relied more, as a conceptual scheme, on hydrodynamics than thermodynamics, both are macroscopic manifestations of the micro-theory of molecular dynamics, thus my thoughts on the kinetic theory approach to quantum gravity \cite{kinQG} and spacetime as condensate \cite{STcond} concepts. I have postponed thinking about the thermodynamics of spacetime because while I find it to be more powerful it is in other ways more restrictive and thus less representative: powerful in what are contained in the laws of thermodynamics, restrictive in the assumption of the equilibrium condition. I feel that if we are to describe the dynamics of spacetime via Einstein equations  \cite{Jac95} or Newton's equation of motion of a particle \cite{Ver10} through thermo- or hydro-dynamics we need to consider nonequilibrium (NEq) conditions ab initio. This explains why I place more emphasis on the nonequilibrium dynamics and thermodynamics for gravitational systems. They are in fact intrinsically nonequilibrium.  Yet of course there is more work devoted to the thermodynamics because we know a great deal more about the equilibrium conditions for the dynamics of both matter and spacetime than the nonequilibrium conditions. Before we end this essay it may therefore be appropriate  to discuss what nonequilbrium means for the problems raised here versus the problem tackled in the thermo-gravity theories of JPV \cite{Jac95,Jac05,Pad10,Ver10}. For matter it is straightforward as can be found in any textbook on nonequilibrium statistical mechanics. For quantum vacuum it is more subtle and tricky -- see d) below.

\textit{a) Gravity is intrinsically non-equilibrium:} the reader probably first encounters this feature from the Jeans
instability. Black hole (equilibrium) thermodynamics is inadequate for treating important issues such as the
end state and information loss issues. Many salient features of gravity stem from the fact that it has negative heat capacity which implies that no stable equilibrium configuration can exist in an isolated gravitational system.  Rather, it requires dynamical considerations (evolution with backreaction) and  nonequilibrium descriptions. As I advocated at the start, it is more natural and productive if we treat and view gravitational phenomena in the idioms and tenets of nonequilibrium physics. What this entails is, understand the general features of NEq physics more thoroughly,  then analyze gravitational systems in that light. We should try to understand better all thermodynamical properties of gravitational systems from this perspective.

\textit{ b) Rely only on the basics:} Instead of tapping into exoteric themes to derive the gravitational equations of motion, we should perhaps first try to invoke only, but delve deeper into,
the nonequilibrium statistical mechanics of gravitating systems.

Indeed let me pose this as a challenge to the young audience here: try to reproduce as many known facts about the thermodynamics of gravitating systems, in particular, black hole thermodynamics,  from the NEq
thermodynamics of systems with long range interactions (LRI).  The part which remains after this filtration is what  truly distinguishes black hole thermodynamics from that of other LRI systems.

\textit{c) Differing philosophies:} In observance of the Austerity Principle and the Commonality Principle
we want to find out what in the familiar black hole thermodynamics cannot be obtained from the more basic and generic features of LRI which places gravity under the same roof with many other physical systems.  This way of thinking has several advantages: 1) Philosophically it is gratifying to see all the implications stemming from one source rather than requiring many ingredients all taken into account with the same degree of importance. What does not fall under these generalities can then be identified as unique about  gravity.  2) To the extent that gravity behaves similar to, say, certain condensed matter systems, the theories describing these more familiar systems can serve as models to discover new phenomena. 3) Experiments related to the analog systems with similar properties can be used to test out predictions in gravitating systems which are difficult to check experimentally. (For example, laboratory cosmology \cite{LabCos}.)

\textit{d) Gravity related to the nonequilibrium matter vs gravity related to the quantum field vacuum: Two distinct levels of inquiry}

What we have studied in this paper is classical matter for which the ordinary concepts and techniques of nonequilibrium statistical mechanics apply. This is very different from the subject of investigation in the recent proposal \cite{Ver10} and the original theory \cite{Jac95} where the nature of gravity in the quantum field vacuum is studied. Concept of equilibrium in that context relies on the Rindler vacuum being thermal with respect to a Minkowski observer, thus invoking the results of Fulling, Davies and Unruh. In \cite{Jac95} the equilibrium considered  is
the local vacuum, viewed from the neighborhood of the bifurcation plane of a local Rindler wedge. No global symmetry or uniform acceleration is invoked there in contrast to \cite{Ver10}, except in the small neighborhood of each point of spacetime. Nonequilibrium refers to the notion of local causal horizons with shear, in which case the system is further from equilibrium than for local horizons without shear. Note also the `system' under consideration in \cite{Jac95}
is defined not as all of the gravitational field, but by the partition created by the local horizon. How does one connect these concepts of temperature and equilibrium associated with a Rindler horizon with the familiar concepts of nonequilibrium statistical mechanics of matter is perhaps the first issue which needs clarification before one can  talk freely about the thermodynamics of spacetime and from there derive equations of motion for matter.

\section*{Appendix: Critique on Gravity as Entropic Force \cite{Ver10}}

Force in polymer chain, crumpling transition. $F \D x= T \D S$.   Ingredients:

1) Bekenstein-Hawking entropy, \textit{holography principle }(screen, etc):
Particle at distance of a Compton wavelength $\D x= h/ mc$ from the BH
horizon would increase the entropy of the BH by one bit $\D S = 2 \pi
k$ (Bekenstein considered a particle fallen into the BH)

2) Relating temperature $T$ to acceleration $a$ via Unruh effect $T =
\hbar a / c k$, get \textit{Newton's Second Law} $F = T (\D S / \D x) = ma$.

3) \textit{For Newton's Law of Gravitation:}  Invoke holography: max number of
bits storage space is proportional to the area of screen: $N = A (c^3 /G \hbar)$.
Relating energy $E$ or mass $M$ of star (enclosed by screen) to $N$ and $T$ by invoking the
equipartition theorem $E = 1/2 NkT= Mc^2$ : counting the number of
degrees of freedom on the event horizon \cite{Pad10}.  Replace $T$ by $E
/ N$ and get $F = T (\D S / \D x) = GMm/ (2 \pi)R^2. ~~ (Area A = 4\pi
R^2 )$ \\

\noindent \textbf{My Critique} -- I'll mention just three for now, one pertaining to the overall intent, two pertaining to central principles:

1) \textit{Why do we need quantum physics from the Compton wavelength of a particle to big ideas like holography or quantum information theory to get mere classical gravity?} This can be brushed off as a matter of preference depending on what ingredients one views as more basic and thus more important. The proponent's argument is that these features come with quantum field theory  and if one can find the root of gravity in these more primitive concepts it is a gain. Preferences can only be persuaded but not be debated. Yet this question gains gravity and becomes more demanding  after what we have seen explained above, i.e, many salient features of gravity are shared by systems with NHC or LRI. Take glassy system for example, do we have to go through this labyrinth of intricate ideas and constructions involving entropy-area, holography or (the equivalent of) Unruh temperature applied to local Rindler observer's vacuum to obtain the equations of motion for the dynamics of ordinary glass? What does one gain in this conceptual maze over the conventional ways in condensed matter physics?

2) \textit{Holography and quantum information}: The area law: entropy $S \sim A$ is a rather generic feature  and has been derived in many ways for various physical systems \cite{Bom86,Sre,PleEis}, not just for black holes. In the most commonly relatable way it can be viewed as  a consequence of partitioning a closed system into subsystems and counting the resultant subsystem's degrees of freedom (e.g., \cite{Pag93}). The projection operator formalism /concepts is quite basic in NEq statistical mechanics. One does not need to invoke big ideas like the holography principle nor new tenets from quantum information to get or appreciate these results.

3) \textit{Thermality of vacuum and dynamics:} Invoking the thermal equilibrium condition for the description of dynamics is, at least by common sense, a bit of a stretch. Arbitrary motion has no thermodynamic description, temperature is ill-defined. Concept of temperature is meaningful only under very special conditions, e.g., Unruh effect manifests only for a detector in uniform acceleration.  General state of motion is under nonequilibrium conditions \cite{RHK} and its full content can only reveal by using  nonequilibrium quantum field theory \cite{CalHu08}. As explained above the notion of equilibrium in the thermo-gravity proposals refers to the local Rindler vacuum which is very different from the notion of equilibrium referring to classical matter or even quantum fields. \\

\noindent {\bf Acknowledgments} I thank the organizers for their invitation to this festive gathering in honour of Mario where I also renewed my special friendship to two of his mighty musketeers, my former collaborators,  Esteban Calzetta and Juan Pablo Paz.  This paper in a more developed form was presented at the Peyresq Physics 15 Meeting funded by the OLAM, Association pour la Recherche Fondamentale, Bruxelles in June 2010 where I enjoyed the warm hospitality of its director, Prof. Edgard Gunzig. I thank Ted Jacobson for explaining his ideas to me once again and valuable critiques on a preliminary draft of this essay, and Werner Israel, Don Page and Rafael Sorkin for their insightful comments. This work is supported in part by NSF grant PHY-0801368.



\begin{thebibliography}{999}

\bibitem{Jac95} T. Jacobson, Phys. Rev. Lett. 75, 1260 (1995).

\bibitem{Jac05} C. Eling, R. Guedens, T. Jacobson, Phys. Rev. Lett. 96,
121301 (2006).


\bibitem{Ver10} E. P. Verlinde,  ``On the Origin of Gravity and the Laws of Newton" [arXiv:1001.0785]

\bibitem{Pad10} T. Padmanabhan, Rep. Prog. Phys. 73 (2010) 046901
``Thermodynamical Aspects of Gravity: New insights" [arXiv:0911.5004].  T. Padmanabhan , Aseem Paranjape, Entropy of Null Surfaces and Dynamics of Spacetime  Phys. Rev. D75 064004, (2007).

\bibitem{GRhydro} B. L. Hu,
"GENERAL RELATIVITY AS GEOMETRO-HYDRODYNAMICS"  Invited talk at the
Second Sakharov International Conference Lebedev Physical Institute,
May, 1996. [gr-qc/9607070].

\bibitem{E/QG} B. L. Hu, ``EMERGENT / QUANTUM GRAVITY: Macro/Micro Structures of Spacetime",
J. Phys. Conf. Ser. 174 (2009) 012015  [arXiv:0903.0878]

\bibitem{Volovik} G. E. Volovik, {\sl The Universe in a Helium Droplet} (Clarendon Press 2003).
``Fermi-point scenario for emergent gravity" in Proceedings of conference "From Quantum to Emergent
Gravity: Theory and Phenomenology" PoS(QG-Ph)043 (2007).

\bibitem{Wen} Xiao-Gang Wen, {\sl Quantum Field Theory of Many-Body Systems} (Oxford University Press 2004).
Michael Levin, Xiao-Gang Wen ``Fermions, strings, and gauge fields in
lattice spin models" Phys. Rev. B67 (2003) 245316, B71, 045110 (2005).

\bibitem{EGrev} B. L. Hu, ``Emergent Gravity: Conceptual Development and New
Challenges" Review for IJMPD (2011)

\bibitem{DICE10} B. L. Hu, ``Gravity and Nonequilibrium Thermodynamics of Quantum Fields" Invited Talk at DICE2010, September 13, 2010, Tuscany, Italy. Proceedings in J. Phys. Conf. Ser. (2011)

\bibitem{Bek72} J.D. Bekenstein, Black holes and entropy, Phys. Rev. D 7, 2333 (1973).

\bibitem{Haw74} S.W. Hawking, Particle creation by black holes, Commun. Math. Phys. 43, 199
(1975).

\bibitem{holography} G. 't Hooft, ``Dimensional reduction in quantum gravity," arXiv:gr-qc/9310026, in Abdus Salam Festschrift: A Collection of Talks (World Scientific, Singapore, 1993).
 L. Susskind, ``The World as a hologram," J. Math. Phys. 36 (1995) 6377 -6396 [arXiv:hep-th/9409089].

\bibitem{Unr76} W.G. Unruh, Notes on black-hole evaporation, Phys. Rev. D 14, 870 (1976).

\bibitem{Austerity} J. A. Wheeler, {\sl Physics and Austerity} (Anhui Science and Technology Publications, Anhui, China, 1982).

\bibitem{Oriti} Daniele Oriti (ed), {\sl Approaches to quantum gravity}  (Cambridge
University Press 2009)

\bibitem{HorPol} Gary T. Horowitz, Joseph Polchinski, ``Gauge/gravity duality" in \cite{Oriti}.


\bibitem{Seiberg} N. Seiberg, ¡°Emergent Spacetime¡± Rapporteur talk at the 23rd Solvay Conference in Physics, December, 2005. [arXiv:hep-th/0601234]

\bibitem{Rovelli} Carlo Rovelli, {\sl Quantum Gravity} (Cambridge University Press, 2004)

\bibitem{Thiemann} Thomas Thiemann, {\sl Modern Canonical Quantum General Relativity}  (Cambridge University Press, 2007)

\bibitem{causet} Invited talks by F. Dowker, R. Sorkin and S. Surya on September 16, 2010 in DICE2010, Tuscany, Italy.
Proceedings in J. Phys. Conf. Ser. (2011).

\bibitem{cosCMP}  B. L. Hu,  ``Cosmology as `Condensed Matter' Physics"  Invited talk given at the Third Asia-Pacific Physics Conference, Hong Kong,  June 1988.  Proceedings edited by K. Young  (World Scientific Publishing Co., Singapore, 1989).)  [gr-qc/9511076]

\bibitem{meso} B. L. Hu, ``Semiclassical Gravity and Mesoscopic Physics"
Invited Talk at the International Symposium on Quantum Classical
Correspondence,   Drexel University, Philadelphia, Sept. 1994,
Proceedings eds D. H. Feng and B. L. Hu   (International Publishers,
Boston, 1997) [gr-qc/9511077].

\bibitem{kinQG} B. L. Hu,  ``A Kinetic Theory Approach to Quantum Gravity" Int. J. Theor.
Phys. 41 (2002) 2111 [gr-qc/0204069]

\bibitem{STcond} B. L. Hu, ``Can Spacetime be a Condensate?"  Int. J. Theor. Phys. 44 (2005)
1785 [gr-qc/0503067]

\bibitem{AdSCFT} J. Maldacena, Adv. Theor. Math. Phys. {\bf 2}, 231 (1998).

\bibitem{Unr81} W.G. Unruh, Phys. Rev. Lett. 46, 1351 (1981)

\bibitem{RHK} A. Raval, B. L. Hu and Don Koks, Phys. Rev. D55,
4795 (1997)

\bibitem{Sak} A. D. Sakharov, ``Vacuum Quantum Fluctuations in Curved Space and the Theory of
Gravitation'' Doklady Akad. Nauk S. S. R. 177, 70-71 (1967) [Sov.
Phys. - Doklady 12, 1040-1041 (1968)].

\bibitem{MTW} C. Misner, K. Thorne and J.A. Wheeler, Gravitation (Freeman, San Francisco,
1972).

\bibitem{LynWoo68} D. Lynden-Bell, R. Wood, Mon. Not. RAS 138 (1968) 495.


\bibitem{HarLafMar} J.B. Hartle, R. Laflamme, and D. Marolf, Phys. Rev. D 51, 7007 (1995).

\bibitem{LynBelPhysica} D. Lynden-Bell, Physica A 263 (1999) 29.

\bibitem{Ant} V.A. Antonov, Vest. Leningrad Gros. Univ. 7 (1962) 135.


\bibitem{PosThi05} H.A. Posch, W. Thirring, Phys. Rev. Lett. 95, 251101
(2005).


\bibitem{Pad} T. Padmanabhan, Phys. Rep. 188 (1990) 285.

\bibitem{Thi70} W. Thirring, Zeitschrift fur Physik 235 (1970) 339.

\bibitem{RLL0TD} A. Ramirez-Hernandez, H. Larralde and F. Leyvraz, Phys. Rev. Lett. 100, 120601 (2008)

\bibitem{Gross_mc}  D.H.E. Gross, {\sl Microcanonical thermodynamics: Phase transitions in ¡°small¡±
systems} (World Scientific, Singapore 2000).

\bibitem{BBY} J. David Brown and James W. York, Jr. ``Microcanonical functional integral for the gravitational field"
Phys. Rev. D 47, 1420 (1993).  Harry W. Braden, J. David Brown, Bernard F. Whiting, and James W. York, Jr.,
Phys. Rev. D 42, 3376 (1990).

\bibitem{NiePRL98} Th. M. Nieuwenhuizen, Phys. Rev. Lett. 81, 2201 (1998)

\bibitem{Siv} C. Sivaram, Phys. Rev. Lett. 84, 3209 (2000).

\bibitem{Campa} A. Camp et al, Phys. Rep. 480, 57 (2009)

\bibitem{HawPag}
S.~W. Hawking and D.~N. Page, Comm. Math. Phys. {\bf 87}, 577 (1983).

\bibitem{Witten} E.~Witten, Adv. Theor. Math. Phys. {\bf 2}, 505 (1998).

\bibitem{HemTho} S. Hemming and L. Thorlacius, JHEP 11 (2007)
086

\bibitem{TayWhe}Edwin F. Taylor and  John Archibald Wheeler, {\sl  Exploring Black Holes: Introduction to General Relativity}  (Addison Wesley Longman, 2000)

\bibitem{OppSpin} J. Oppenheim, Phys. Rev. E68, 016108 (2003)

\bibitem{PVI98} F. Pretorius, D. Vollick and W. Israel, Phys. Rev. D57, 6311
(1998).

\bibitem{OppArea} J. Oppenheim, Phys. Rev. D65, 024020 (2001)

\bibitem{OliDam} H. P. de Oliveira and I. Damiao Soares, Phys. Rev. D71, 124034 (2005)

\bibitem{GravColcridyn} M.W. Choptuik, Phys. Rev. Lett. 70, 9 (1993).
A. M. Abrahams and C. R. Evans, Phys. Rev. Lett. 70, 2980 (1993).


\bibitem{Penrose} R. Penrose, in S. Hawking and G. Ellis (eds), Einstein
Centenary Volume, (Cambridge University Press, 1979)

\bibitem{Hu81} B. L. Hu, 
Phys. Lett. 97A, 368 (1983)

\bibitem{Liberati} G. Chirco, S. Liberati, Phys. Rev. D81, 024016 (2010)

\bibitem{LebLie} J.L. Lebowitz, E.H. Lieb, Phys. Rev. Lett. 22 (1969) 613

\bibitem{PosThi06} H.A. Posch, W. Thirring, Phys. Rev. E 74 (2006) 051103.

\bibitem{LabCos} 
E. Calzetta and B. L. Hu, Int. J. Theor. Phys. 44 (2005) 1691	[cond-mat/0503367]

\bibitem{Bom86} L. Bombelli, R. K. Koul, J. Lee, and R. D. Sorkin, Phys.
Rev. D 34, 373 (1986)

\bibitem{Sre} M. Srednicki, Phys. Rev. Lett. 71, 666 (1993)

\bibitem{PleEis} J. Eisert, M. Cramer and M. B. Plenio, Rev. Mod. Phys. 82, 277 (2010)

\bibitem{Pag93} D.N. Page, Phys. Rev. Lett. 71, 1291 (1993).

\bibitem{CalHu08}  E. Calzetta and B. L. Hu, {\sl Nonequilibrium Quantum Field Theory}
(Cambridge University Press, 2008)

\end{thebibliography}
\end{document}